\documentstyle[12pt]{article}
\def\beq{\begin{equation}}
\def\eeq{\end{equation}}
\def\bea{\begin{eqnarray}}
\def\eea{\end{eqnarray}}
\def\bef{\begin{figure}}
\def\enf{\end{figure}}
\def\S{{\bf S}}
\def\C{{\bf C}}
\def\Z{{\bf Z}}
\def\R{{\bf R}}
\def\N{{\bf N}}

\def\P{{\bf P}}
\def\CC{{\cal C}}
\def\CF{{\cal F}}

\def\CN{{\cal N}}
\def\CO{{\cal O}}

\def\We{{W_{\mbox{eff}}}}

\def\ba{\begin{array}}
\def\ea{\end{array}}
\def\bce{\begin{center}}
\def\ece{\end{center}}

\def\La{\Lambda}

\def\ev#1{\langle#1\rangle}
\def\vol#1{{\bf #1}}

\def\nuphb#1#2#3{Nucl. Phys. {\bf B}\vol{#1#2#3} }

\begin{document}
\begin{titlepage}
 \rightline{hep-th/0411013} \vskip 1cm \centerline{
\Large\bf{Open-Closed String Duality from Orbifolded Conifolds}}
\vskip 1cm

\centerline{\sc Seungjoon Hyun$^{a}$ and Kyungho Oh$^{b}$ } \vskip
1cm  \centerline{$^a$ Institute of Physics and Applied Physics, }
\centerline{ Yonsei University, Seoul 120-749, Korea }
\centerline{\tt } \centerline{ $^b$ Department of Mathematics and
Department of Physics and Astronomy,}  \centerline{University of
Missouri-St. Louis, St. Louis, MO 63121, USA }  \vskip 2cm
\centerline{\sc Abstract} We study large N dualities for a class
of ${\cal N} = 1$ theories realized on type IIB D5 branes wrapping
2-cycles of local Calabi-Yau threefolds which is obtained from
resolving orbifolded conifolds or as effective field theories on
D4 branes in type IIA brane configurations. The field theory is
$\CN =1$ supersymmetric $\prod U(N_{ij})$ Yang-Mills gauge theory.
Strong coupling effects are analyzed in the deformed geometry. We
propose open-closed string duality via a geometric transition in
toric geometry.  The T-dual type IIA picture and M-theory lifting
are also considered.
\def\today{\ifcase\month\or
January\or February\or March\or April\or May\or June\or
July\or August\or September\or October\or November\or December\fi,
\number\year}
\vskip 1cm
\end{titlepage}
\newpage
\section{Introduction}
\setcounter{equation}{0} Duality between open and closed string is
one of the most intriguing features of string theory and impetus
for current activities in topological string theory~\cite{top1,
top2,top3}. The large $N$ duality between Chern-Simons theory on
the $\bf S^3$ cycle of the deformed conifold and topological
closed strings on the resolved conifold was observed in
\cite{gova}  and the result was embedded in type IIA strings by
Vafa \cite{vafa}. The topological transition becomes a geometric
transition between D6 branes on the $\bf S^3$ of the deformed
conifold and type IIA strings on the resolved conifold, with
fluxes or between D5 branes wrapped on the $\P^1$ cycle of the
resolved conifold and type IIB on the deformed conifold with
fluxes. The type IIB formulation has been extended to a large
class of geometric transition dualities, for various local
Calabi-Yau
geometries\cite{GT,sv,civ,eot,ckv,GUD,dot1,dot2,Klebanov,KW}.

In the present paper, we begin with orbifolded conifolds $\C_{kl}$
which is an orbifold of the three dimensional conifold $xy -uv =0$
by a discrete group $\Z_k \times \Z_l$. The resulting singularity
is a toric singularity whose equation can be written after
coordinate changes as \bea \label{orbconifold} xy =z^k, uv
=z^l\eea
 and we describe a
partial resolution of the singularity where the worst
singularities are conifolds. Then we take a small resolution of
this geometry by replacing each singular point by $\P^1$. The
normal bundle of $\P^1 $ will be $\CO (-1) + \CO(-1)$ and by
wrapping $N_{ij}$ D5 branes around each $\P^1$ cycles, we obtain
$\CN =1$ supersymmetric gauge theory with gauge group
$\prod_{i=1}^k \prod_{j=1}^l U(N_{ij})$ from type IIB open string
theory. Following AdS/CFT, the branes will disappear while fluxes
remain through three Lagrangian cycles in a new geometry in the
large $N_{ij}$ limit. Now in the strong coupling regime,  each
gauge group has a gluino condensation. This new geometry is
obtained by deforming the partially resolved orbifolded conifold
where $\P^1$ cycles are shrunken and are replaced by the $\S^3$
cycles with RR, NS fluxes. The gluino condensation are mapped into
the sizes of $\S^3$ cycles which are periods of holomorphic three
forms.

The $\C^*$ action on the orbifolded conifold by \bea \lambda \cdot
(x,y,u,v,z) \rightarrow (\lambda x, \lambda^{-1} y, \lambda u,
\lambda^{-1}v, z), \lambda \in \C^* \eea on (\ref{orbconifold})
naturally extends to the resolved orbifolded conifold and by
taking T-dual along the $U(1)$ direction, one obtain type IIA
pictures which have been explored in a series of
papers~\cite{dot1,dot2,dot3}. We may now lift this type IIA theory
to eleven dimensional M theory where the brane configuration give
rises to a Riemann surface embedded in a complex three dimensional
space.

\section{Vafa's open-closed string dualities}

We will briefly review some of features of  Vafa's $\CN =1$ open
and closed string  dualities for large N which will be used later.
Consider type IIB theory on a non-compact Calabi-Yau threefold
$O(-1) + O(-1)$ of $\P^1$ which is a small resolution of the
conifold : \bea xy -uv =0 \eea by wrapping $N$ D5 branes on
$\P^1$. This gives a four dimensional $\CN=1$ $U(N)$ pure
Yang-Mills theory described by open strings ending on the D5
branes, in the small 't Hooft parameter regime. Vafa's duality
states that in the large $N$ limit (large 't Hooft parameter
regime), this is equivalent to type IIB on the deformed conifold:
\bea f= xy -uv -\mu =0. \eea In the deformed conifold, the $\P^1$
cycle is shrunken and replaced by $\S^3$ of size $\mu$ which is
identified with the condensate of the $SU(N)$ glueball superfield
$S = -\frac{1}{32 \pi^2} \mbox{Tr} W_\alpha W^\alpha$. The
description is now in terms of closed strings. Rather than the $N$
original D5 branes, there are now $N$ units of RR flux through
$\S^3$, and also some NS flux through the non-compact cycle dual
to $\S^3$. The glueball $S$ is identified with the flux of the
holomorphic 3-form on the compact 3-cycle of the deformed conifold
\bea S = \int_A \Omega \eea and the integral of  holomorphic
3-form on the noncompact 3-cycle is made with introducing a
cut-off $\La_0$: \bea \Pi = \int_B^{\La_0} \Omega = \frac{1}{2\pi
i}(-3S \log \La_0 -S + S \log S)\eea The effective superpotential
is written as \bea W_{\mbox{eff}} = \int (H_{RR} + \tau H_{NS})
\wedge \Omega \eea where $H_{RR}$ is the RR flux on the A-cycle
and is due to the $N$ D5 branes and  $H_{NS}$ is the NS flux on
the noncompact B-cycle. By using the usual IR/UV identification in
the AdS/CFT conjecture, we identify the large distance  $\La_0$
(small IR scale) in supergravity with the small distance (large UV
scale) in the field theory such that the coupling constant in
field theory is constant and finite in UV. After doing so, the
form of the effective superpotential is \bea W_{\mbox{eff}} = S
\log [\La^{3N}/S^N] +NS \eea The condition of supersymmetry
implies that the derivative of $ W_{\mbox{eff}}$ with respect to
$S$ is zero which implies that $S$ gets $N$ discrete values,
separated by a phase. This is the gluino condensation in the field
theory and signals the breaking of the chiral symmetry ${\bf Z_{2
N} \rightarrow Z_2}$. The gluons of $SU(N)$ get a mass so the
$SU(N)$ gets a mass gap and confines. What remains is the $U(1)$
part of $U(N)$ whose coupling constant is equal to the coupling
constant of the $U(N)$ theory divided by $N$.

The lift of the transition to M theory by using MQCD was
considered in \cite{dot1,dot2,dot3}. A T-duality of the
geometrical constructions takes the N D5 branes wrapped on $\P^1$
to a brane configuration with two orthogonal NS branes on the
directions $x$ and $y$ (together with four directions
corresponding to the Minkowski space) and $N$ D4 branes in the the
direction $x_n$. The lift to M theory involves a single M5 brane
which has the worldvolume $R^{1,3} \times \Sigma$ where $\Sigma$
is a 2-dim. manifold holomorphically embedded in $(x, y, t)$ where
$t = exp(\frac{x_n}{R_{10}} + i~x_{10})$. When the $\P^1$ cycle
shrinks, the direction $x_n$ goes to zero and eventually the
coordinate $t$ becomes the coordinate of a circle. Because we
cannot embed holomorphically into a circle, it results that the
coordinate $t$ of the M5 brane become constant and $\Sigma$ is
embedded inside $x~y~=~\mbox{const.} $ where the constant is
related to the scale of the $U(N)$ theory. After reducing to ten
dimensions,  $x~y~=~\mbox{const.} $ becomes the equations for a
2-dimensional surface where an NS brane is wrapped, which is
T-dual to the deformed conifold and the constant is related to the
size of the $\S^3$ cycle. Therefore we could explicitly see the
relation between the scale of the $U(N)$ theory and the size of
the $\S^3$ cycle.

The transition has been generalized to more complicated geometries
in \cite{sv,civ,eot,ckv,GUD}, where the blown-up geometry involves
more
 $\P^1$ cycles and the deformed geometry involves more  $\S^3$ cycles.
In \cite{GT, GUD}, it has been explored that  the large N
dualities from type IIB, type IIA and M-theory perspectives for
large classes of the $\CN=1$ supersymmetric gauge theories by
wrapping D5 branes on blown-up $\P^1$ cycles in Calabi-Yau
threefolds which are obtained by deforming resolved ALE spaces of
the ADE singularities. It begins with the  ${\cal N} = 2$ quiver
gauge associated with the ADE Dynkin diagrams, which can be
geometrically engineered on the resolved ALE spaces. Then the
$\CN=1$ theory is obtained by adding a superpotential for the
adjoint fields, $W_i(\Phi_i)$, thus the full tree-level
superpotential is \bea \label{SUPER} W =\sum_{i=1}^n W_i
-\mbox{Tr}\sum_{i=1}^{n}\sum_{j=1}^n s_{i,j}Q_{i,j} \Phi_{j}
{Q}_{j,i},~~\mbox{where }  W_i = \mbox{Tr}\sum_{j=1}^{d_i+1}
g_{i,j-1} \Phi_i^j, \eea with adjoints $\Phi$ and bifundamentals
$Q_{i,j}$, which fixes the moduli space of the ${\cal N} = 1$
theory.

\section{Toric Description of Orbifolded Conifolds }
In this section, we begin with toric description of orbifolded
conifolds and the D-brane world-volume gauge theory on its
resolution. The orbifolded conifolds have been studied in various
aspects \cite{dhot, otorb}. A toric variety is a space which
contains algebraic torus ${(\C^*)}^d$ as an
 open dense subset. For example, a projective space
${\bf P}^d =  (\C^{d+1} -\{0\})/\C^*$ is a toric variety because it contains
$(\C^*)^d \cong (\C^*)^{d+1}/\C^* \subset (\C^{d+1} -\{0\})/\C^*$.
As in the case of the projective space, we will express our toric varieties as
a quotient space (this can be thought of as a holomorphic quotient in
the sense of the Geometric Invariant Theory~\cite{mfk}
 or as
a symplectic reduction as in gauged linear sigma model. In our cases, these two
will be the same \cite{kir}.):
\bea
V_\Delta = (\C^q - F_\Delta)//(\C^*)^{q-d}
\eea
where $q, F_\Delta$ and the action of $(\C^*)^{q-d}$ on $C^q$ are determined
by a combinatorial data $\Delta$.
Now we give a description of the combinatorial data $\Delta$
for  Gorenstein canonical
singularity (i.e. a singularity with a trivial canonical class, $K$).
Consider  vectors  $v_1, \ldots , v_q$ in a lattice $\N = \Z^d \subset
\N_{\R}=\N \otimes \R =\R^d$ in general position.
We introduce the corresponding homogeneous coordinates $x_i$ for
of  $\C^q- F_\Delta$ in the holomorphic quotients. In gauged linear sigma
model, these correspond to  matter multiplets.
 There will be $(q-d)$ independent relations
\bea
\sum_{i=1}^q Q_i^{(a)}v_i = 0, \quad a=1, \ldots , q-d.
\eea
Here $Q^{(a)}$'s correspond to the charges of the matter fields under
$U(1)^{q-d}$ which is the maximal compact subgroup of $(\C^*)^{q-d}$.
The D-term equations will be
\bea
\sum_{i=1}^{q} Q^{(a)}_i |x_i|^2 = r_a, \quad a =1, \ldots , q-d.
\eea
In the holomorphic quotient, the charge matrix whose column vectors
consist of  $Q^{(a)}$
determines the  action of $(\C^*)^{q-d}$ on $\C^q$ i.e. the action of
$(\lambda_1, \lambda_2, \ldots , \lambda_{q-d}) \in (\C^*)^{q-d}$ on
$(x_1, \ldots , x_q) \in \C^q$ is given by
\bea
(\lambda_1^{Q_1^{(1)}}
\lambda_2^{Q_1^{(2)}}
\cdots
\lambda_{q-d}^{Q_1^{(q-d)}}x_1,
\lambda_1^{Q_2^{(1)}}
\lambda_2^{Q_2^{(2)}}\cdots
\lambda_{q-d}^{Q_2^{(q-d)}}x_2,
\ldots ,
\lambda_1^{Q_q^{(1)}}
\lambda_2^{Q_q^{(2)}}
\cdots
\lambda_{q-d}^{Q_q^{(q-d)}}x_q)
\eea
Here the action can be carried out
as written or in two steps, an $(\R_+)^{q-d}$ action and a $U(1)^{q-d}$ action
if K\"ahler. The quotient will depend on the gauge fixing determined by
the $(\R_+)^{q-d}$ action i.e. the moment map. In the holomorphic approach,
this corresponds to different spaces $F_\Delta$
which give rise to (partial) resolutions of
the original space $V_\Delta$.
In toric diagram, this corresponds to different
triangulations of a convex cone in $\R^d$ determined
by $\{ v_1, \ldots , v_q\}$. The collection of these combinatorial
data is  denoted by  $\Delta$ called a fan.
The quotient space $V_\Delta$
will have Gorenstein canonical singularity if there exists
$u \in \Z^d$ such that $u \cdot v_i =1$ for all $i$ \cite{reid}.
Thus $v_i$'s will
lie on the hyperplane with normal $u$ at a distance $1/\|u \|$
from the origin in $\R^d$.
This imposes the following condition on the charge vectors $Q^{(a)}$:
\bea
\sum_{i=1}^q Q^{(a)}_i = 0, \quad a=1, \ldots , q-d
\eea

To put our discussions in the language of the gauged linear sigma model,
recall that $\C^q$ is a symplectic manifold with the standard symplectic
form $\omega = i\sum_{i=1}^q dz^i \wedge d\bar{z}^{\bar{i}}$. The maximal
compact subgroup $G:=U(1)^{q-d}$ of $(\C^*)^{q-d}$ acts covariantly
on a symplectic manifold $(\C^q, \omega)$ by symplectomorphisms.
The infinitesimal action will give rise to a moment map $\mu: \C^q \to
{\bf g}^*$ by Poisson brackets.
In coordinates, the components of $\mu : \C^q \to \R^{q-d}$ are given by
\bea
\label{moment}
\mu_a = \sum_{i=1}^q Q^{(a)}_i |x_i|^2 - r_a
\eea
where $r_a$ are undetermined additive constants. The symplectic reduction
is then defined as
\bea
V(r) \equiv \mu^{-1}(0)/G.
\eea
The structure of $V(r)$ will depend on $r$. It follows from (\ref{moment})
that every $(\C^*)^{q-d}$-orbit in $\C^q$ will contribute at most one point
to $V(r)$. The value of $r$ will determine
the contributing orbits. For a fixed  $r$,
the set of $(\C^*)^{q-d}$-orbits which do not contribute is precisely
$F_\Delta$. The quotient space $V(r)$ carries a symplectic form $\omega_r$
by reducing $\omega$. The symplectic reduction carries a natural complex
structure, in which the reduced symplectic form becomes a K\"ahler form.

Now we will consider quotient singularities of the conifold (i.e. orbifolded
conifold). The conifold is a three dimensional hypersurface singularity in
$\C^4$ defined by:
\bea
{\cal C}: \quad xy -uv = 0.
\eea
The conifold can be realized as a holomorphic quotient of $\C^4$
by the $\C^*$ action given by \cite{witten, KW}
\bea
(A_1, A_2,B_1, B_2)\mapsto (\lambda A_1, \lambda A_2,\lambda^{-1} B_1,
\lambda^{-1} B_2)\quad\mbox{ for }\lambda \in \C^*.
\eea
Thus the charge matrix is the transpose of $Q^{'}
=(1,1,-1,-1)$ and $\Delta=\sigma$ will be a convex polyhedral cone
in $\N^{'}_{\R}=\R^3$
generated by $v_1, v_2, v_3, v_4 \in \N^{'}=\Z^3$  where
\bea
v_1=(1,0,0), \quad v_2=(0,1,0),\quad  v_3=(0,0,1),\quad
v_4=(1,1,-1).
\eea
The isomorphism between the conifold ${\cal C}$ and the holomorphic
quotient is given by
\bea
\label{act}
x=A_1B_1, \quad y=A_2B_2, \quad u=A_1B_2, \quad v=A_2B_1.
\eea
We take a further quotient of the conifold ${\cal C}$ by a discrete group
$\Z_k \times \Z_l$. Here $\Z_k$ acts on $A_i, B_j$ by
\bea
\label{zk}
(A_1, A_2, B_1, B_2) \mapsto
(e^{-2\pi i/k} A_1, A_2, e^{2\pi i/k}B_1, B_2),
\eea
and $\Z_l$ acts by
\bea
\label{zl}
(A_1, A_2, B_1, B_2) \mapsto
(e^{-2\pi i/l} A_1, A_2, B_1, e^{2\pi i/l}B_2).
\eea
Thus they will act on the conifold ${\cal C}$ by
\bea
\label{xy}
(x,y,u,v) \mapsto (x,y,e^{-2\pi i/k}u, e^{2\pi i/k}v)
\eea
and
\bea
\label{uv}
(x,y,u,v) \mapsto (e^{-2\pi i/l}x,e^{2\pi i/l}y, u, v).
\eea
Its quotient is  called the hyper-quotient of the conifold
or the orbifolded conifold and denoted by ${\cal C}_{kl}$.
To put the actions (\ref{act}), (\ref{zk}) and (\ref{zl}) on an equal footing,
consider the over-lattice $\N$:
\bea
\N = \N^{'} + \frac{1}{k}(v_3-v_1) + \frac{1}{l}(v_4 -v_1).
\eea
Now the lattice  points $\sigma \cap \N$ of $\sigma$ in $\N$
 is generated by
$(k+1)(l+1)$ lattice points as a semigroup (These lattice points will be
referred as a toric diagram.). The charge matrix $Q$  will be
$(k+1)(l+1)$ by $(k+1)(l+1)-3$. The discrete group $\Z_k \times \Z_l \cong
\N / \N^{'}$ will act on the conifold $\C^4 // U(1)$ and its quotient
will be the symplectic reduction $\C^{(k+1)(l+1)} // U(1)^{(k+1)(l+1)-3}$
with the moment map associated with  the charge matrix $Q$. The new toric
diagram for ${\cal C}_{kl}$ will also lie
on the plane at a distance $1/\sqrt{3}$ from the origin with a normal vector
$(1,1,1)$ and we draw a toric diagram on the plane for ${\cal C}_{57}$:
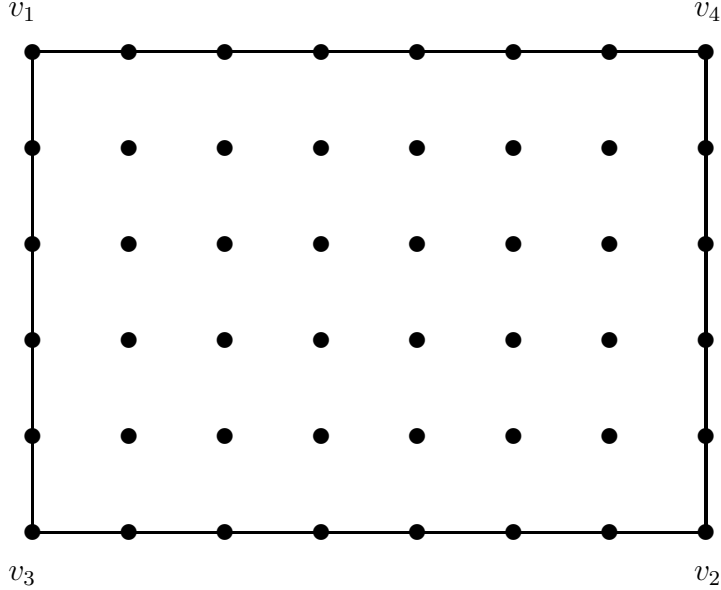
\begin{figure}
\setlength{\unitlength}{0.00083300in}%
\begingroup\makeatletter\ifx\SetFigFont\undefined%
\gdef\SetFigFont#1#2#3#4#5{%
  \reset@font\fontsize{#1}{#2pt}%
  \fontfamily{#3}\fontseries{#4}\fontshape{#5}%
  \selectfont}%
\fi\endgroup%
\begin{picture}(1000,3705)(2000,-4897)
\thicklines
\put(3601,-2161){\circle*{100}}
\put(3601,-2761){\circle*{100}}
\put(4201,-1561){\circle*{100}}
\put(3601,-3361){\circle*{100}}
\put(3601,-3961){\circle*{100}}
\put(4801,-1561){\circle*{100}}
\put(5401,-1561){\circle*{100}}
\put(6001,-1561){\circle*{100}}
\put(6601,-1561){\circle*{100}}
\put(7201,-1561){\circle*{100}}
\put(7201,-2161){\circle*{100}}
\put(7201,-2761){\circle*{100}}
\put(7201,-3361){\circle*{100}}
\put(7201,-3961){\circle*{100}}
\put(6601,-3961){\circle*{100}}
\put(6001,-3961){\circle*{100}}
\put(5401,-3961){\circle*{100}}
\put(4801,-3961){\circle*{100}}
\put(4201,-3961){\circle*{100}}
\put(4201,-3361){\circle*{100}}
\put(4801,-3361){\circle*{100}}
\put(5401,-3361){\circle*{100}}
\put(6001,-3361){\circle*{100}}
\put(6601,-3361){\circle*{100}}
\put(6601,-2761){\circle*{100}}
\put(6601,-2161){\circle*{100}}
\put(6001,-2161){\circle*{100}}
\put(6001,-2761){\circle*{100}}
\put(5401,-2161){\circle*{100}}
\put(5401,-2761){\circle*{100}}
\put(4801,-2161){\circle*{100}}
\put(4801,-2761){\circle*{100}}
\put(4201,-2161){\circle*{100}}
\put(4201,-2761){\circle*{100}}
\put(7801,-1561){\circle*{100}}
\put(7801,-2161){\circle*{100}}
\put(7801,-2761){\circle*{100}}
\put(7801,-3361){\circle*{100}}
\put(7801,-3961){\circle*{100}}
\put(3601,-1561){\circle*{100}}
\put(3601,-4561){\circle*{100}}
\put(4201,-4561){\circle*{100}}
\put(4801,-4561){\circle*{100}}
\put(5401,-4561){\circle*{100}}
\put(6001,-4561){\circle*{100}}
\put(6601,-4561){\circle*{100}}
\put(7201,-4561){\circle*{100}}
\put(7801,-4561){\circle*{100}}
\put(3451,-1336){$v_1$}
\put(7726,-1336){$v_4$}
\put(3451,-4861){$v_3$}
\put(7726,-4861){$v_2$}
\put(3601,-1561){\line( 0,-1){3000}}
\put(3601,-4561){\line( 1, 0){4200}}
\put(7801,-4561){\line( 0, 1){3000}}
\put(7801,-1561){\line(-1, 0){4200}}
\end{picture}

\caption{A toric diagram for ${\bf Z}_5\times {\bf Z}_7$ hyper-quotient
of the conifold, ${\cal C}_{57}$}
\end{figure}
The action $(\ref{xy}), (\ref{uv})$ of $\Z_k \times \Z_l$ on the
conifold ${\cal C}$ can be lifted to an action on $\C^4$ whose
coordinates are $x,y, u, v$. The ring of invariants will be
$\C[x^l, y^l, xy, u^k, v^k, uv]$ and the orbifolded conifold
${\cal C}_{kl}$ will be defined by the ideal $(xy-uv)\C[x^l, y^l,
xy, u^k, v^k, uv]$. Thus after renaming variables, the defining
equation for the orbifolded conifold will be \bea \label{con-eqn}
{\cal C}_{kl}: xy =z^l, \quad uv =z^k. \eea Figure 1 shows the
toric diagram for ${\cal C}_{57}$.
\section{Open String Description on Small resolution of orbifolded conifolds}
We resolve the orbifolded conifold in two steps. First let us
consider a partial resolution of the orbifolded conifold where all
singularities are conifold singularities. This is obtained by
subdivision of the toric diagram. This subdivision is done by
adding edges among the adjacent lattice points in the toric
diagram i.e. the points in $\sigma \cap \N$. Now there are $kl$
singular points and all of them are conifold singularities. The
new toric diagram for this partial resolution for ${\cal C}_{57}$
is shown in Figure 2.
\begin{figure}
\setlength{\unitlength}{0.00083300in}%
\begingroup\makeatletter\ifx\SetFigFont\undefined%
\gdef\SetFigFont#1#2#3#4#5{%
  \reset@font\fontsize{#1}{#2pt}%
  \fontfamily{#3}\fontseries{#4}\fontshape{#5}%
  \selectfont}%
\fi\endgroup%
\begin{picture}(1000,3705)(2000,-4897)
\thicklines
\put(3601,-1561){\circle*{100}}
\put(3601,-2161){\circle*{100}}
\put(3601,-2761){\circle*{100}}
\put(3601,-3361){\circle*{100}}
\put(3601,-3961){\circle*{100}}
\put(3601,-4561){\circle*{100}}

\put(4201,-1561){\circle*{100}}
 \put(4801,-1561){\circle*{100}}
\put(5401,-1561){\circle*{100}} \put(6001,-1561){\circle*{100}}
\put(6601,-1561){\circle*{100}} \put(7201,-1561){\circle*{100}}
\put(7201,-2161){\circle*{100}} \put(7201,-2761){\circle*{100}}
\put(7201,-3361){\circle*{100}} \put(7201,-3961){\circle*{100}}
\put(6601,-3961){\circle*{100}} \put(6001,-3961){\circle*{100}}
\put(5401,-3961){\circle*{100}} \put(4801,-3961){\circle*{100}}
\put(4201,-3961){\circle*{100}} \put(4201,-3361){\circle*{100}}
\put(4801,-3361){\circle*{100}} \put(5401,-3361){\circle*{100}}
\put(6001,-3361){\circle*{100}} \put(6601,-3361){\circle*{100}}
\put(6601,-2761){\circle*{100}} \put(6601,-2161){\circle*{100}}
\put(6001,-2161){\circle*{100}} \put(6001,-2761){\circle*{100}}
\put(5401,-2161){\circle*{100}} \put(5401,-2761){\circle*{100}}
\put(4801,-2161){\circle*{100}} \put(4801,-2761){\circle*{100}}
\put(4201,-2161){\circle*{100}} \put(4201,-2761){\circle*{100}}
\put(7801,-1561){\circle*{100}} \put(7801,-2161){\circle*{100}}
\put(7801,-2761){\circle*{100}} \put(7801,-3361){\circle*{100}}
\put(7801,-3961){\circle*{100}}

\put(4201,-4561){\circle*{100}} \put(4801,-4561){\circle*{100}}
\put(5401,-4561){\circle*{100}} \put(6001,-4561){\circle*{100}}
\put(6601,-4561){\circle*{100}} \put(7201,-4561){\circle*{100}}
\put(7801,-4561){\circle*{100}} \put(3451,-1336){$v_1$}
\put(7726,-1336){$v_4$} \put(3451,-4861){$v_3$}
\put(7726,-4861){$v_2$} \put(3601,-1561){\line( 0,-1){3000}}
\put(3601,-4561){\line( 1, 0){4200}} \put(3601,-1561){\line( 1,
0){4200}} \put(3601,-2161){\line( 1, 0){4200}}
\put(3601,-2761){\line( 1, 0){4200}} \put(3601,-3361){\line( 1,
0){4200}} \put(3601,-3961){\line( 1, 0){4200}}
\put(3601,-4561){\line( 1, 0){4200}}

\put(3601,-1561){\line(0,-1){3000}}
\put(4201,-1561){\line(0,-1){3000}}
\put(4801,-1561){\line(0,-1){3000}}
\put(5401,-1561){\line(0,-1){3000}}
\put(6001,-1561){\line(0,-1){3000}}
\put(6601,-1561){\line(0,-1){3000}}

\put(7201,-1561){\line(0,-1){3000}}

 \put(7801,-4561){\line(0,1){3000}}

\put(7801,-1561){\line(-1, 0){4200}}

\put(7801,-4561){\line(0,1){3000}}

\put(7801,-1561){\line(-1, 0){4200}}

\end{picture}

\caption{A toric diagram for a partially resolved orbifolded
conifold, ${\cal C}_{57}$}
\end{figure}
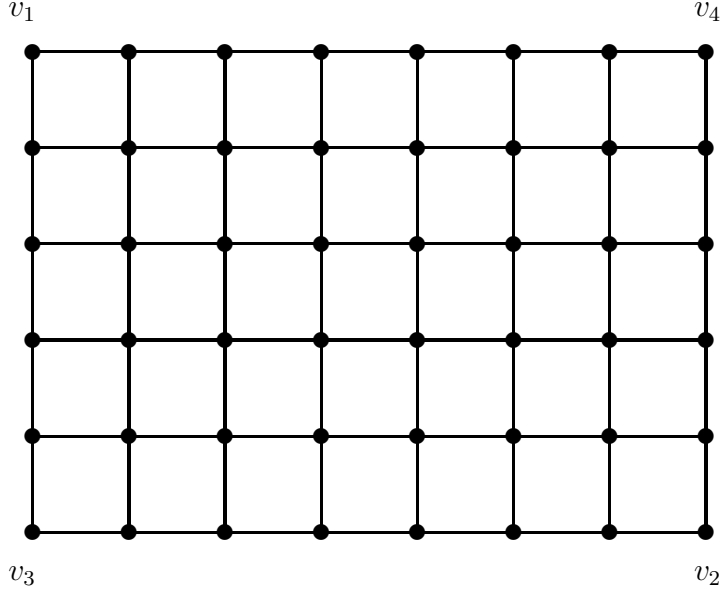
Now we make a further subdivision of the toric diagram by putting
edges between the vertices in the diagonal direction. For each
square, we put only one edge and we have two choices i.e.
northeast or northwest diagonals. The resulting geometry for this
choice are related by flops and they induces Seiberg dualites as
explained in \cite{GT,dot2} in the gauge theory we are about to
explain. This will create  $kl$ new $\P^1$'s in the geometry and
the local geometry around each $\P^1$'s can be identified with
$\CO (-1) + \CO(-1)$ of $\P^1$. The resulting geometry will be
called a resolved orbifolded conifold or a small resolution of the
orbifolded conifold.  By putting $N_{ij}$ D5 branes on each
$\P^1$, we consider type IIB theory on the world volume of
D-branes. This gives a four dimensional $\CN =1$ pure Yang-Mills
theory described by strings ending on the D5 branes with gauge
group \bea \prod_{i=1}^{k}\prod_{j=1}^{l}U(N_{ij}). \eea If the
chain of $\P^1$'s where D-branes are wrapped on is disconnected,
then there will be no interaction among  gauge theories of each
connected component in low energy limit. So we will assume that
the chain of $\P^1$'s where D-branes are wrapped on is connected.
Then there are only two choices for this kind of small resolution.
Figure 3 shows one of them for ${\cal C}_{57}$ and the other one
is obtained by changing the direction of diagonal edges.
\begin{figure}
\setlength{\unitlength}{0.00083300in}%
\begingroup\makeatletter\ifx\SetFigFont\undefined%
\gdef\SetFigFont#1#2#3#4#5{%
  \reset@font\fontsize{#1}{#2pt}%
  \fontfamily{#3}\fontseries{#4}\fontshape{#5}%
  \selectfont}%
\fi\endgroup%
\begin{picture}(1000,3705)(2000,-4897)
\thicklines \put(3601,-1561){\circle*{100}}
\put(3601,-2161){\circle*{100}} \put(3601,-2761){\circle*{100}}
\put(3601,-3361){\circle*{100}} \put(3601,-3961){\circle*{100}}
\put(3601,-4561){\circle*{100}}

\put(4201,-1561){\circle*{100}}
 \put(4801,-1561){\circle*{100}}
\put(5401,-1561){\circle*{100}} \put(6001,-1561){\circle*{100}}
\put(6601,-1561){\circle*{100}} \put(7201,-1561){\circle*{100}}
\put(7201,-2161){\circle*{100}} \put(7201,-2761){\circle*{100}}
\put(7201,-3361){\circle*{100}} \put(7201,-3961){\circle*{100}}
\put(6601,-3961){\circle*{100}} \put(6001,-3961){\circle*{100}}
\put(5401,-3961){\circle*{100}} \put(4801,-3961){\circle*{100}}
\put(4201,-3961){\circle*{100}} \put(4201,-3361){\circle*{100}}
\put(4801,-3361){\circle*{100}} \put(5401,-3361){\circle*{100}}
\put(6001,-3361){\circle*{100}} \put(6601,-3361){\circle*{100}}
\put(6601,-2761){\circle*{100}} \put(6601,-2161){\circle*{100}}
\put(6001,-2161){\circle*{100}} \put(6001,-2761){\circle*{100}}
\put(5401,-2161){\circle*{100}} \put(5401,-2761){\circle*{100}}
\put(4801,-2161){\circle*{100}} \put(4801,-2761){\circle*{100}}
\put(4201,-2161){\circle*{100}} \put(4201,-2761){\circle*{100}}
\put(7801,-1561){\circle*{100}} \put(7801,-2161){\circle*{100}}
\put(7801,-2761){\circle*{100}} \put(7801,-3361){\circle*{100}}
\put(7801,-3961){\circle*{100}}

\put(4201,-4561){\circle*{100}} \put(4801,-4561){\circle*{100}}
\put(5401,-4561){\circle*{100}} \put(6001,-4561){\circle*{100}}
\put(6601,-4561){\circle*{100}} \put(7201,-4561){\circle*{100}}
\put(7801,-4561){\circle*{100}} \put(3451,-1336){$v_1$}
\put(7726,-1336){$v_4$} \put(3451,-4861){$v_3$}
\put(7726,-4861){$v_2$} \put(3601,-1561){\line( 0,-1){3000}}
\put(3601,-4561){\line( 1, 0){4200}}
\put(3601,-1561){\line( 1,
0){4200}} \put(3601,-2161){\line( 1, 0){4200}}
\put(3601,-2761){\line( 1, 0){4200}} \put(3601,-3361){\line( 1,
0){4200}} \put(3601,-3961){\line( 1, 0){4200}}
\put(3601,-4561){\line( 1, 0){4200}}

\put(3601,-1561){\line(0,-1){3000}}
\put(4201,-1561){\line(0,-1){3000}}
\put(4801,-1561){\line(0,-1){3000}}
\put(5401,-1561){\line(0,-1){3000}}
\put(6001,-1561){\line(0,-1){3000}}
\put(6601,-1561){\line(0,-1){3000}}

\put(7201,-1561){\line(0,-1){3000}}

\put(7801,-4561){\line(0,1){3000}}

\put(7801,-1561){\line(-1, 0){4200}}

\put(3601,-2161){\line(1,1 ){600}}

\put(3601,-2161){\line( 1, -1){2400}}

\put(3601,-3361){\line( 1, 1){1800}}

\put(3601,-3361){\line( 1, -1){1200}}

\put(3601,-4561){\line( 1, 1){3000}}
\put(7801,-1561){\line(-1,-1){3000}}

\put(7801,-2761){\line(-1,-1){1800}}
\put(7801,-2761){\line(-1,1){1200}}
\put(7801,-3961){\line(-1,-1){600}}
\put(7801,-3961){\line(-1,1){2400}}
\put(7201,-4561){\line(-1,1){3000}}
\end{picture}

\caption{A toric diagram for  small resolution of the orbifolded
conifold, ${\cal C}_{57}$}
\end{figure}
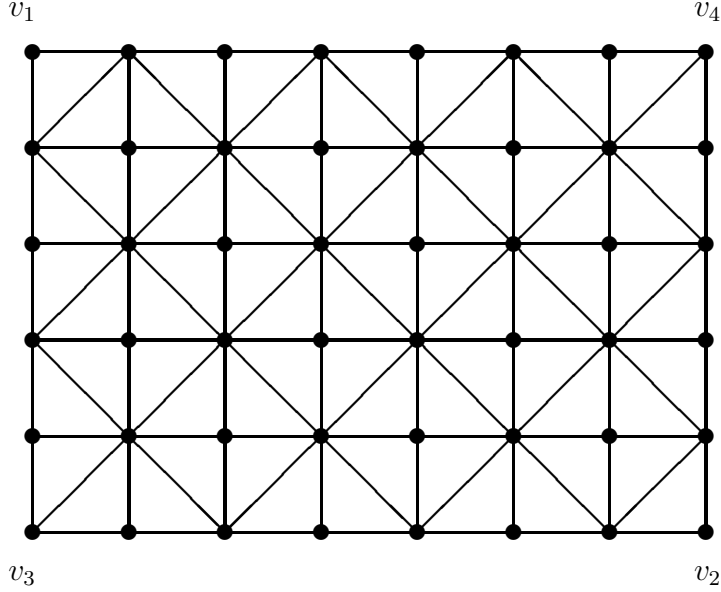
To relate this to type IIA theory, consider a circle action on the
orbifolded conifold \bea {\cal C}_{kl}: xy =z^l, \quad uv =z^k.
\nonumber \eea given by \bea \label{sc}x \rightarrow e^{i\theta}
x, y \rightarrow e^{-i\theta} y, u \rightarrow e^{i\theta} u, v
\rightarrow e^{-i\theta} v \eea Then this action lifts to an
action on the small resolution of orbifolded conifold and the
lifted action acts on the normal bundle $\CO (-1) + \CO(-1)$ of
$\P^1$. To describe the action, we introduce two copies of $\C^3$
with coordinates $Z, X, Y$ (resp. $Z', X', Y'$) for the first
(resp. second) $\C^3$. Then $\CO(-1) + \CO (-1)$ over $\P^1$ is
obtained by gluing two copies of $\C^3$ with the identification:
\bea \label{-1-1} Z' = \frac{1}{Z} ~, ~\quad X'= XZ ~, ~\quad Y' =
YZ~. \eea The $Z$ (resp. $Z'$) is a coordinate of $\P^1$ in the
first (resp. second) $\C^3$ and others are the coordinates of the
fiber directions. The blown-down map from the resolved conifold
$\C^3 \cup \C^3$ to the conifold $\CC$ is given by \bea x=X=X'Z',
~~y=ZY=Y',~~u = ZX = X', ~~v= Y = Z'Y'. \eea From this map, one
can see that the following action  on the resolved conifold is an
extension of the action (\ref{sc}): \bea (e^{i \theta }, Z) \to
e^{i \theta} Z, ~~(e^{i\theta}, X) \to X,~~ (e^{i\theta},
Y) \to e^{-i\theta}Y\nonumber \\
 (e^{i \theta }, Z') \to e^{-i \theta} Z', ~~(e^{i\theta}, X') \to
e^{i\theta}X',~~ (e^{i\theta}, Y') \to Y' \label{sk} \eea The
orbits degenerates along the union of two complex lines $Z=Y=0$ in
the first copy of $\C^3$ and $Z'=Y'=0$  in the second copy of
$\C^3$. Note that these two lines do not intersect and in fact
they are separated by the size of $\P^1$. Now we take T-dual along
the orbits of $S_r$ of type IIB theory obtained by wrapping $N$ D5
branes on the rigid $\P^1$. Again there will be two NS branes
along the degeneracy loci of the action: one NS brane, denoted by
$NS_X$, spaced along $X$ direction (which is defined  by $Z=Y=0$
in the first $\C^3$)
 and the other NS brane, denoted by $NS_{Y'}$  along
$Y'$ direction (which is defined by $Z'=X'=0$ in the second
$\C^3$). Therefore the T-dual picture will be a brane
configuration of D4 brane along the interval with two NS branes in
the `orthogonal' direction at the ends of the the interval. Here
the length of the interval is the same as the size of the rigid
$\P^1$.  Hence the type IIA brane picture of the orbifolded
conifold will be a web of D4 branes in the interval and NS branes
on the intersections. Figure 4 shows the type IIA brane
configuration of  ${\cal C}_{57}$ corresponding to type IIB
picture Figure 3 where D4 branes are along the edges and NS branes
are located at the dots.
\begin{figure}
\setlength{\unitlength}{0.00083300in}%
\begingroup\makeatletter\ifx\SetFigFont\undefined%
\gdef\SetFigFont#1#2#3#4#5{%
  \reset@font\fontsize{#1}{#2pt}%
  \fontfamily{#3}\fontseries{#4}\fontshape{#5}%
  \selectfont}%
\fi\endgroup%
\begin{picture}(1000,3705)(2000,-4897)
\thicklines 
\put(3601,-2161){\circle*{100}} 
\put(3601,-3361){\circle*{100}} 
\put(3601,-4561){\circle*{100}}

\put(4201,-1561){\circle*{100}}
\put(5401,-1561){\circle*{100}} 
\put(6601,-1561){\circle*{100}} 
\put(7201,-2161){\circle*{100}} 
\put(7201,-3361){\circle*{100}} 
\put(6601,-3961){\circle*{100}} 
\put(5401,-3961){\circle*{100}} 
\put(4201,-3961){\circle*{100}} 
\put(4801,-3361){\circle*{100}} 
\put(6001,-3361){\circle*{100}} 
\put(6601,-2761){\circle*{100}}
\put(6001,-2161){\circle*{100}} 
\put(5401,-2761){\circle*{100}}
\put(4801,-2161){\circle*{100}} 
\put(4201,-2761){\circle*{100}}
\put(7801,-1561){\circle*{100}} 
\put(7801,-2761){\circle*{100}} 
\put(7801,-3961){\circle*{100}}

\put(4801,-4561){\circle*{100}}
\put(6001,-4561){\circle*{100}}
\put(7201,-4561){\circle*{100}}
%
%
%
%

\put(3601,-2161){\line(1,1 ){600}}

\put(3601,-2161){\line( 1, -1){2400}}

\put(3601,-3361){\line( 1, 1){1800}}

\put(3601,-3361){\line( 1, -1){1200}}

\put(3601,-4561){\line( 1, 1){3000}}
\put(7801,-1561){\line(-1,-1){3000}}

\put(7801,-2761){\line(-1,-1){1800}}
\put(7801,-2761){\line(-1,1){1200}}
\put(7801,-3961){\line(-1,-1){600}}
\put(7801,-3961){\line(-1,1){2400}}
\put(7201,-4561){\line(-1,1){3000}}
\end{picture}

\caption{Type IIA brane picture of  small resolution of the
orbifolded conifold, ${\cal C}_{57}$}
\end{figure}
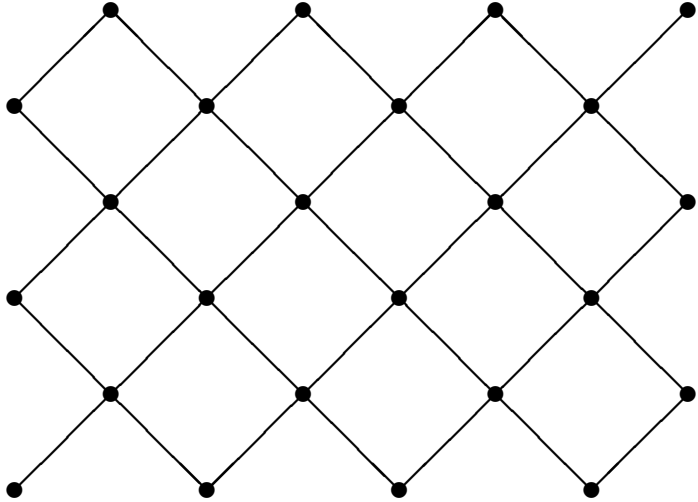
\section{Closed String Description on Deformed Orbifolded Conifold}
In the large $N_{ij}$ limit in large 't Hooft parameter regime, we
have gaugino condensation and mass gap as noticed in \cite{ckv}.
The theory lives on a geometry where the $\P^1$ cycles have shrunk
and $\S^3$ cycles have grown and RR fluxes through them and NS
fluxes through their dual cycles have been created. This new
geometry is a complex deformation of the partially resolved
orbifolded conifold.  As one may have noticed, we are only
deforming a part of the orbifolded conifold. These are so called
normalizable deformations. These normalizable deformations
correspond to dynamical fields.  The local geometry around $\P^1$
cycle where $N_{ij}$ D branes wrapped goes into a geometric
transition to a deformed conifold and  the deformed geometry is
given by an equation \bea x^2 + y^2 + u^2 + v^2 = \zeta_{ij}. \eea
with holomorphic three form \bea \Omega \sim \frac{dx\, dy\,
du}{\sqrt{\zeta_{ij} - x^2 -y^2 -u^2}}. \eea Now $\P^1$
disappeared and three sphere $A_{ij} \cong\S^3$ has been created
instead. The three sphere lies as a Lagrangian 3-cycle in the
local geometry which is the cotangent bundle of $\S^3$. The period
of the holomorphic three-form $\Omega$ over the 3-cycle $A_{ij}$
is given by \bea S_{ij} = \int_{A_{ij}} \Omega \sim
\frac{\zeta_{ij}}{4}. \eea The period over the non-compact dual
 cycle $B_{ij}$ is divergent and hence we introduce a cut off
$\La_{0}$ to regulate the infinity: \bea \Pi_{ij}
=\int_{B_{ij}}^{\La_{0}} \Omega \sim \frac{1}{2\pi i} \left(
-3S_{ij} \log \La_{0} - S_{ij} + S_{ij} \log S_{ij}\right) +
\ldots. \eea As in the other geometric transitions ~\cite{civ,
ckv}, $S_{ij}$ is identified with the glueball field $S_{ij} =
-\frac{1}{32 \pi^2} \mbox{Tr}_{SU(N_{ij})} W_\alpha W^\alpha $ of
the non-Abelian factor $SU(N_{ij})$  of $U(N_{ij})$ in the dual
theory. The $S_{ij}$ will be massive and obtain particular
expectation values due to the superpotential $\We$. The dual
superpotential $\We$ arises from the non-zero fluxes left after
the transition. The deformed geometry will have $N_{ij}$ units of
$H_R$ flux through the $A_{ij} \cong \S^3$ cycle due to $N_{ij}$
D5 branes wrapped on the $\P^1$ cycle before the transition, and
there is also an $H_{NS}$ flux $\alpha_{ij}$ through each of the
dual non-compact cycle $B_{ij}\cong \R^3$ with $2\pi i \alpha_{ij}
= 8\pi^2/g_0^2$ given in terms of the bare coupling constant $g_0$
of the $U(N_{ij})$ theory.

Thus the effective superpotential is \bea\label{supoteff} -
\frac{1}{2\pi i} \We &=&\sum_{i=1}^k\sum_{j=1}^l \left(
\int_{A_{ij}} H_R \int_{B_{ij}} \Omega -\int_{B_{ij}} H_{NS}
\int_{A_{ij}}\Omega \right)\nonumber\\ &=&\sum_{i=1}^k\sum_{j=1}^l
(N_{ij} \Pi_{ij} + \alpha_{ij} S_{ij}). \eea After identifying
$\La_{0}$ with the UV-cutoff $\La$ in the dual gauge theory, we obtain the usual low energy
superpotential associated with the $SU(N_{ij})$ glueballs: \bea
\We = \sum_{i=1}^k\sum_{j=1}^l S_{ij}\left(\log \frac{\La^{3
N_{ij}}}{S_{ij}^{N_{ij}}} + {N_{ij}}\right). \eea Integrating out
the massive $S_{ij}$ by solving \bea \frac{\partial \We}{\partial
S_{ij}} =0 \eea leads to $N_{ij}$ supersymmetric vacua of
$SU(N_{ij})$ $\CN =1$ supersymmetric Yang-Mills: \bea \label{Sij}
\ev{S_{ij}} = \exp (2\pi i m/N_{ij}) \La^3, ~~m=1, \ldots,
N_{ij}. \eea

The dual theory obtained after the transition is an $\CN =2$
\bea\prod_{i=1}^k\prod_{j=1}^l U(1) \equiv U(1)^{kl}\eea gauge
theory broken to $\CN=1$ $U(1)^{kl}$ by the superpotential $\We$
(\ref{supoteff})~\cite{tv}. The $S_{ij}$, which are the same
$\CN=2$ multiplet as the the $U(1)^{kl}$, get masses and frozen to
particular $\ev {S_i}$ by $\We$. On the other hand, the $\CN=1$
$U(1)^{kl}$ gauge fields remain massless. The couplings
$\tau_{iji'j'}$ of these $U(1)$'s can be determined by $\Pi_{ij}$
or the $\CN =2$ prepotential $\CF(S_{ij})$, with $\Pi_{ij} =
\partial \CF /\partial S_{ij}$ : \bea \tau_{iji'j'} =
\frac{\partial \Pi_{ij}}{\partial S_{i'j'}} = \frac{\partial^2 \CF
(S_{ij})}{\partial S_{ij}\partial S_{i'j'}}.\eea The couplings
should be evaluated at the vacua $\ev {S_{ij}}$ obtained in
(\ref{Sij}). It is possible to consider M thoery by lifting type
IIA brane configuration which is constructed in the previous
section and investigate the large N limit via Witten's MQCD
formalism. We denote the finite direction of D4 branes by $x^7$
and the angular coordinate of the circle $\S^1$ in the 11-th
dimension
 by $x^{10}$. Thus the NS branes are separated along the $x^7$
direction. We combine them into a complex coordinate \bea t = \exp
( -R^{-1}x^7 - i x^{10})\eea where $R$ is the radius of the circle
$\S^1$ in the 11-th dimension. In MQCD \cite{witqcd}, the
classical type IIA brane configuration turns into a single
fivebrane whose world-volume is a product of the Minkowski space
$\R^{1,3}$ and the M-theory curve $\Sigma$ in a flat Calabi-Yau
manifold \bea \label{M}M = \C^2 \times \C^*,\eea whose coordinates
are $u,v,t$. Classically , the $u, v$-coordinates of the M-theory
curve $\Sigma$ describes the location of NS branes in the type IIA
picture.  By lifting this to the M-theory, the location of the D4
branes on the NS brane will be smeared out as D4 brane acquire
11-th direction. So, at quantum level, the M-theory curve is
obtained by merging tubular neighborhood around edges in Figure 4
which represents D4 branes and NS branes located dots in Figure 4.
A lifting from the compact part of the Jacobian of the curves to
the non-compact part can be viewed as a geometric transition from
M-theory point of view~\cite{dot1}.


\section{Acknowledgments}
We would like to thank Jae-Suk Park, Jong-Dae Park and Sang-Heon Yi for inspiring discussions and
encouragement. This work  was supported by Brain Pool program No. 031S-1-6 from The Korean Federation of Science and Technology Societies. K.O. thanks IPAP at Yonsei University for warm hospitality during his visit.

\end{document}